\documentstyle[12pt]{article}
\begin{document}

\input epsf

\begin{titlepage}

\begin{flushright}
Freiburg--THEP 97--08\\
March 1997
\end{flushright}
\vspace{2.5cm}
\begin{center}

\large\bf

{\LARGE\bf Radiative corrections to the lineshape 
           of a heavy Higgs boson at muon colliders}\\[2cm]
\rm
{Thomas Binoth and Adrian Ghinculov}\\[.5cm]

{\em Albert--Ludwigs--Universit\"{a}t Freiburg,
           Fakult\"{a}t f\"{u}r Physik}\\
      {\em Hermann--Herder Str.3, D-79104 Freiburg, Germany}\\[3.5cm]
      
\end{center}
\normalsize

\begin{abstract}
We study the NNLO radiative corrections of enhanced electroweak 
strength to the lineshape of the Higgs boson in the processes 
$\mu^+\mu^-\rightarrow H\rightarrow ZZ$ and
$\mu^+\mu^-\rightarrow H\rightarrow t\bar t$,
which are of interest for Higgs searches at a possible muon collider.
We give the radiative corrections both in the on--shell renormalization 
scheme and in the pole renormalization scheme. The pole scheme
appears to have better convergence properties, in agreement with previous
results regarding the position of the Higgs pole in the complex plane.
\end{abstract}

\end{titlepage}


\title{Radiative corrections to the lineshape 
       of a heavy Higgs boson at muon colliders}

\author{Thomas Binoth and Adrian Ghinculov}

\date{{\em Albert--Ludwigs--Universit\"{a}t Freiburg,
           Fakult\"{a}t f\"{u}r Physik},\\
      {\em Hermann--Herder Str.3, D-79104 Freiburg, Germany}}

\maketitle

\begin{abstract}
We study the NNLO radiative corrections of enhanced electroweak 
strength to the lineshape of the Higgs boson in the processes 
$\mu^+\mu^-\rightarrow H\rightarrow ZZ$ and
$\mu^+\mu^-\rightarrow H\rightarrow t\bar t$,
which are of interest for Higgs searches at a possible muon collider.
We give the radiative corrections both in the on--shell renormalization 
scheme and in the pole renormalization scheme. The pole scheme
appears to have better convergence properties, in agreement with previous
results regarding the position of the Higgs pole in the complex plane.
\end{abstract}


As higher loop calculation are becoming available in 
the electroweak symmetry breaking sector of the standard model,
one is able to perform phenomenological studies of processes 
relevant for Higgs searches at future colliders at higher orders
in perturbation theory.
The precision of such studies is restricted by the renormalization scheme 
dependence which enters through the truncation
of the perturbative series of a given observable.

The scheme dependence is formally one order higher
than the order calculated. This is a small effect
as long as the the coupling constant is small.
As the coupling constant increases,
the sizes of different loop order corrections are growing,
the scheme dependence of the predictions increases, and
the perturbative result is affected by large theoretical uncertainties.
By comparing several schemes with respect to the size 
of the loop corrections, one can get some insight on the reliabilty
of the perturbative calculation in these schemes. One is especially interested
in schemes which have better convergence properties.
Because the mass of the Higgs boson is related
to the quartic coupling of the Higgs field,
such effects must be taken into account
when performing phenomenological studies related to
the search for a heavy Higgs boson.

In ref. \cite{higgspole} we studied the position of the pole
of the Higgs propagator at three--loop level.
There we concluded that it appears to be more convenient to describe
heavy Higgs bosons in terms of the pole renormalization scheme 
instead of the widely used on--shell scheme.
We suggested that this may be the result of the presence
in the Higgs sector of a nonperturbative 
mechanism similar to the effect discussed in ref. \cite{beenakker}
for the case of a simpler model which admits a solution at all orders.

In this letter we study the NNLO radiative corrections for 
the reactions $\mu^+\mu^-\rightarrow H \rightarrow ZZ$
and $\mu^+\mu^-\rightarrow H \rightarrow,t\bar t$. These processes 
were already studied at leading order as a source of Higgs bosons at
a possible muon collider \cite{gunion}.
We give the radiative corrections both
in the on--shell and the pole renormalization schemes, and
compare the two schemes.
The results support the previous conclusion about the convergence of 
perturbative expansions in the pole scheme versus the on--shell scheme.

Let us start with the on--shell scheme, in which most of the
existing one-- and two--loop results were derived.
In the following we will calculate the ratios between the 
amplitudes of the processes considered, including the radiative
corrections up to the desired order, and the corresponding amplitudes
in the Born approximation:

\begin{eqnarray}\label{kdefos}
  K_1(s) & = & V_{H\mu^+\mu^-} \,
             \frac{s-m_H^2+i \,m_H \Gamma_{tree}}{s-m_H^2+i \,m_H \Pi(s)} \, 
	       V_{Ht\bar{t}}   \nonumber \\
  K_2(s) & = & V_{H\mu^+\mu^-} \,
             \frac{s-m_H^2+i \,m_H \Gamma_{tree}}{s-m_H^2+i \,m_H \Pi(s)} \, 
	       V_{HZZ}(s)   
\end{eqnarray}
These factors can easily be included in a Monte Carlo simulation
for taking into account the NNLO radiative corrections.

In the equations above, $\Pi(s)$ is the Higgs self--energy, and in the
resonance region plays essentially the role of an $s$ dependent width.
$V_{H\mu^+\mu^-} \equiv V_{Ht\bar{t}}$ and $V_{HZZ}(s)$ are the radiative 
correction factors to the Yukawa couplings, and to the Higgs coupling 
to the weak vector bosons, respectively. $\Pi$ and $V_{HZZ}$
are functions of $s$, while the corrections to the Yukawa couplings are
momentum--independent at leading order in the Higgs mass 
\cite{2loop:Htott,kniehl}.

The existing one-- and two--loop results in the Higgs sector 
allow one to determine the correction factors $K_1$ and $K_2$ at
NNLO. 
The vertex corrections are known up to ${\cal O}(\lambda^2)$, and
the Higgs selfenergy is known to ${\cal O}(\lambda^3)$.
In the resonance region the instability of the Higgs boson
needs to be taken into acoount by means of a Dyson summation.
This introduces inverse powers of the coupling $\lambda$.
At the same time, in order to keep all contributions relevant 
at the order considered, one must remember that in the resonance 
region the quantity $\delta = (s-m_H^2)/m_H^2$ is of order $\lambda$.
Therefore one has to perform a double expansion in $\delta$ 
and $\lambda$ of the quantities entering the expressions of $K_1$
and $K_2$ \cite{higgspole,glufusion}:

\begin{eqnarray}
\Pi(s)/m_H &=& A + \delta B + \delta^2 C 
                 + {\cal O}(\delta^3\lambda) \nonumber\\
&&\nonumber \\
A &=& a_1\lambda + a_2 \lambda^2 + a_3 \lambda^3 + {\cal O}(\lambda^4) \nonumber\\
B &=& b_1\lambda + b_2 \lambda^2 + {\cal O}(\lambda^3) \nonumber\\
C &=& -i c_1 \lambda + {\cal O}(\lambda^2) \nonumber\\
&&\nonumber \\
V_{H\mu\mu,Ht\bar t} &=& 1 + d_1\lambda + d_2 \lambda^2 
                           + {\cal O}(\lambda^3)\nonumber\\
V_{Hzz}(s) &=& 1 + e_1\lambda + e_2 \lambda^2 + f_1 \delta\lambda
                + {\cal O}(\lambda^3)
\end{eqnarray}

The necessary coefficients $a_i$---$f_i$ can be extracted from 
one-- and two--loop calculations already available
\cite{marciano}--\cite{jikia}:

\begin{eqnarray}
&& a_1 = 3\pi/8 \, ,\quad a_2 = a_1\cdot0.350119 \,,
\quad  a_3 = a_1\cdot (0.97103+0.000476)\,, \nonumber \\
&& b_1 = 0 \,,\quad b_2 = 1.002245 \,,\quad c_1 = 0.2181005\, , \nonumber \\
&& d_1 = 0.132325 \,,\quad d_2 = -0.26387\, ,\nonumber \\
&&e_1 = 0.17505951 - i\, 1.4190989\,,\quad e_2 = -0.53673- i\, 0.32811\,, \nonumber \\
&&f_1 = 0.42536605 - i\, 0.15169744\quad . \nonumber
\end{eqnarray} 

These constants define the correction factors $K_1$ and $K_2$
around the Higgs resonance correct at NNLO in the on--shell 
renormalization scheme (OS).
In this scheme, the Higgs mass $m_H$
is defined by the zero of the real part of the inverse Higgs
propagator. In the pole scheme (PS)
the mass $M$ of the Higgs boson is defined by
the real part of the location of the pole in the complex $s$ plane.

To relate to the pole renormalization scheme, one needs 
the relation between the on--shell mass $m_H$ and the pole mass $M$.
Such a relation, correct at order $\lambda^3$, 
was derived in refs. \cite{valencia,higgspole}:

\begin{equation}\label{massrel}
  M^2 = m_H^2 \left[ 1 + \frac{a_1^2}{4} 
              \left(\frac{G_F}{\sqrt{8}\pi^2} \right)^2 
  m_H^4 + \frac{a_1 a_2 - 2 a_1 b_2 + 2 a_1^2 c_1 }{2}  
  \left(\frac{G_F}{\sqrt{8}\pi^2} \right)^3 m_H^6 \right] 
\end{equation}
This implicitly gives the relation between the Higgs quartic couplings
$\lambda      = G_F m_H^2/(\sqrt{8}\pi^2)$ and
$\lambda^{PS} = G_F   M^2/(\sqrt{8}\pi^2)$, defined in the two schemes.
$G_F = 1.16637 \cdot 10^{-5} \; GeV^{-2}$ is the Fermi constant.
It is straightforward to use eq. 3 for translating 
the expansions of eqns. 2 in the pole renormalization scheme,
and thus obtain the correction factors $K_1$ and $K_2$ in the pole
scheme. Of course, the tree level widths are given by 
$\Gamma^{OS}_{tree}=a_1 m_H \lambda$, and 
$\Gamma^{PS}_{tree}=a_1 M \lambda^{PS}$.

We plot the correction factors $K_1$ and $K_2$ in the on--shell
and the pole schemes in figs. 1 and 2, 
for several values of the Higgs mass.
In order to compare the two renormalization schemes, 
one has to fix the relative values
of the Higgs masses in the two schemes, in a way which 
would correspond in a certain sense to the same physical situation.  
This is an ambiguous choice. If we relate the two masses 
by eq. \ref{massrel}, this is equivalent to fix
the real part of the position of the pole of the Higgs propagator 
at the same place. This is the convention we used in figs. 1 and 2.
Other relations are of course possible. For example, an equally
justified choice is to compare the corrections for values
of the coupling constants for which the Higgs width is the same.
Another possibility is to use the position of the peak of a given process 
to relate the coupling constants of the two schemes.
Obviously, this choice is process dependent.

The factors $K_1$ and $K_2$, which are plotted in figs. 1 and 2, 
can easily be included in a Monte Carlo simulation in order to
take into account the NNLO radiative corrections of enhanced 
electroweak strength to the Higgs lineshape. 
One only needs to multiply the tree 
level amplitude by the factor corresponding to the desired 
renormalization scheme.
One notes that our expressions are only valid around the Higgs
peak. Far from the resonance there is no reason to perform 
the Dyson summation, and one needs the exact self--energy
and vertex functions instead of the momentum expansion which we use.

The size of the NNLO correction, compared to the NLO correction,
is smaller in the pole scheme than in the on--shell
scheme for the $\mu^+\mu^-\rightarrow H \rightarrow ZZ$
and $\mu^+\mu^-\rightarrow H \rightarrow,t\bar t$ processes
in the resonance region. 
This behaviour can be seen clearly in fig. 3, where we plot the absolute
value of the correction factors $K_1$ and $K_2$. It agrees with
similar results concerning the position of the Higgs pole in the
complex plane \cite{higgspole}, and it may be the result of a certain
nonperturbative effect in the Higgs sector of the standard model
\cite{higgspole,beenakker}. 

We regard this behaviour as an indication
that the perturbative expansion converges better in the 
pole scheme than in the on--shell scheme, making the pole scheme
preferable for studying the phenomenology of heavy Higgs bosons. 
This argument was discussed 
in some detail in refs. \cite{higgspole,q96,2loop:Htoww,2loop:Htott}, 
and we will not repeat it here. 
It relies on the assumption that the behaviour 
of the first few terms in the perturbative expansion is indicative
for the divergence pattern of the asymptotic series. Of course,
it is difficult to establish unambiguously if that is indeed 
the case without explicit knowledge of higher order corrections.
Some insight in the behavoiur of higher orders can be obtained
indirectly by studying the scale dependence of the results
within some class of renormalization schemes, like for instance
$\overline{MS}$ \cite{riesselmann,willenbrock}.
Considering the complexity of the existing two--loop calculations
in the Higgs sector, there is little hope that such physical processes
will become available at three--loop order in the near future.
One only has an incomplete picture of the Higgs sector in the 
strong coupling zone, and all existing information should be
kept in mind.

\begin{figure}
\hspace{1.5cm}
    \epsfxsize = 13cm
    \epsffile{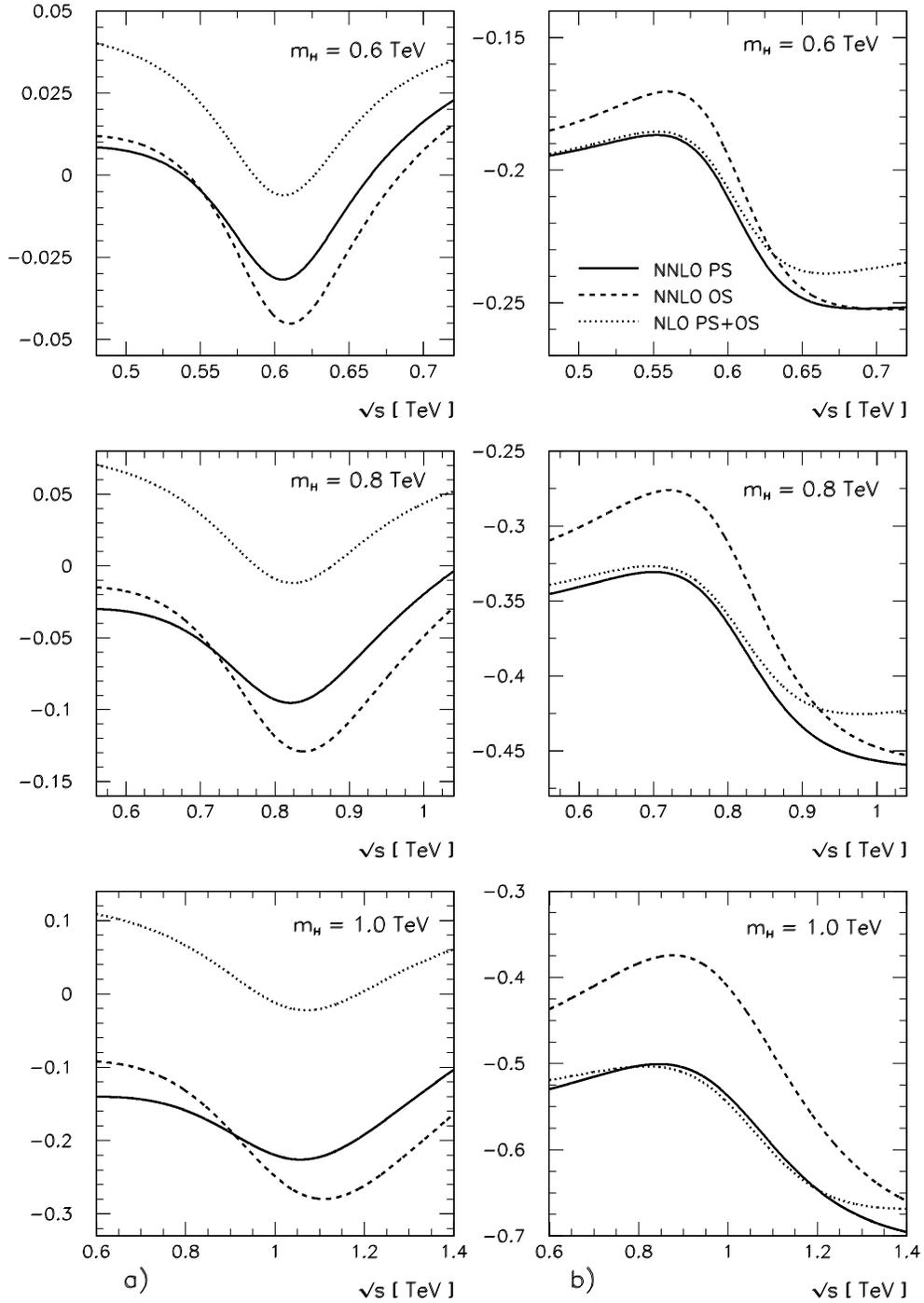}
\caption{{\em The real (a) and the imaginary part (b) of 
              the radiative correction factor to the
              process $\mu^+\mu^- \rightarrow H \rightarrow ZZ$,
              for different values of the on--shell mass
              parameter $m_H$, which correspond to pole masses 
	      as defined in the text. We plot in this picture the
	      value of $K_1-1$.}}
\end{figure}

\begin{figure}
\hspace{1.5cm}
    \epsfxsize = 13cm
    \epsffile{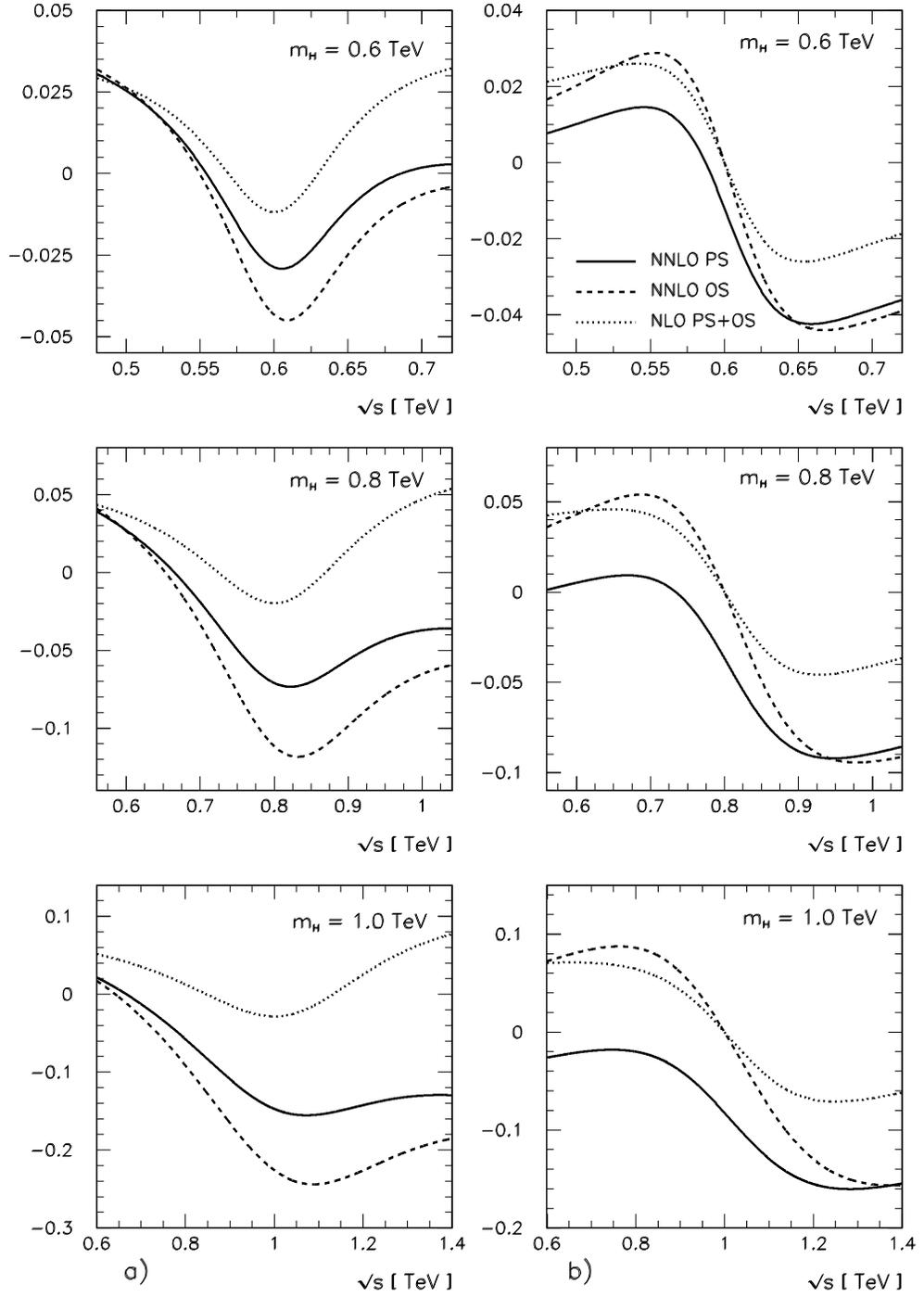}
\caption{{\em Same as fig. 1, but for the correction factor 
              $K_2$ of the 
              process $\mu^+\mu^- \rightarrow H \rightarrow t\bar{t}$.}}
\end{figure}

\begin{figure}
\hspace{1.5cm}
    \epsfxsize = 13cm
    \epsffile{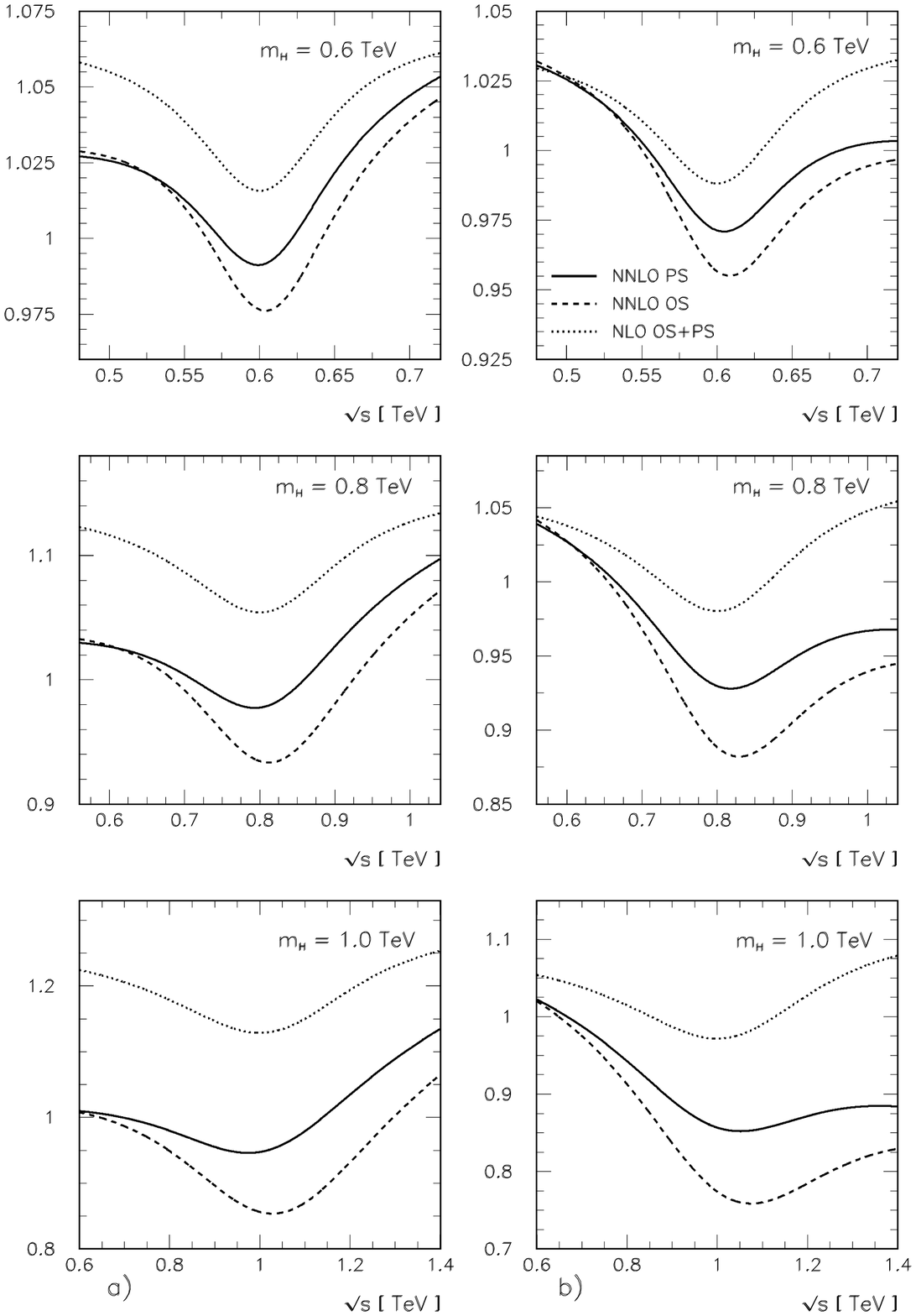}
\caption{{\em The absolute values of the radiative correction
              factors $K_1$ (a) and $K_2$ (b) for different values
              of the quartic coupling. We compare the pole
              and the on--shell schemes at values of the Higgs
              masses related as explained in the text.}}
\end{figure}   


\vspace{.5cm}

{\bf Acknowledgements}

We are indebted to Scott Willenbrock, Jochum van der Bij,
George Jikia and Boris Kastening
for interesting discussions.
The work of A. G. is supported 
by the Deutsche Forschungsgemeinschaft (DFG).


\end{document}